\documentclass[twocolumn,showpacs,amsmath,amssymb]{revtex4}

\usepackage[dvips]{graphicx}
\usepackage{graphics}
\usepackage{dcolumn}
\usepackage{bm}

\begin{document}

\title{Momentum Distribution Function of a Narrow Hall Bar in
the FQHE Regime}

\author{S.-R. Eric Yang$^{1}$, Sami Mitra$^{2}$, M.P.A. Fisher$^{3}$,
and A.H. MacDonald$^{2}$}
\address{$^{1}$Department of Physics, Korea University, 
Seoul 136-701, Korea\\
$^{2}$Department of Physics,
Indiana University, Bloomington, Indiana 47405\\
$^{3}$Institute for Theoretical Physics, University of California, 
Santa Barbara, California 93106}

\begin{abstract}
The momentum distribution function ($n(k)$) of a narrow 
Hall bar in the fractional quantum Hall effect regime is investigated
using Luttinger liquid and microscopic many-particle wavefunction
approaches.  For wide Hall bars with filling factor $\nu=1/M$, where $M$ 
is an odd integer, $n(k)$ has singularities at 
$\pm M k_F$.  We find that for narrow Hall bars additional 
singularities occur at smaller odd integral multiples of $k_F$:
$ n(k) \sim A_p \mid k\pm pk_{F} \mid ^{2\Delta_{p}-1}$
near $k=\pm pk_{F}$, where $p$ is an odd integer $M,M-2,M-4,...,1$.
If inter-edge interactions can be neglected, the 
exponent $2 \Delta_{p}= (1/ \nu+p^{2} \nu)/2$ is independent of the 
width ($w$) of the Hall bar but the amplitude of the singularity $A_p$ 
vanishes exponentially with $w$ for $p\not=M$. 
\end{abstract}
\maketitle


Incompressible states associated with the fractional quantum 
Hall effect have gapless edge excitations which can be 
described using a chiral Luttinger liquid model. 
At $\nu =1/M$ the momentum distribution functions
of these liquids have singular contributions  
proportional to $ |k \pm Mk_{F}|^{M-1}$ 
where $k_{F}$ is the Fermi wavevector\cite{W,KF,MM,HR}.
The singularities give rise to a number of power laws for
the dependence of observables on temperature or on 
voltage which differ from the power laws which would 
apply for Fermi liquids and can, at least in principle,
be tested experimentally. 
Initial experiments\cite{MUW} appear to confirm the Luttinger liquid model
predictions.  Experimental systems in which 
Luttinger liquid model predictions are expected to be tested,
often involve the formation of electrostatically defined constrictions 
in the two-dimensional electron gas, which bring opposite sides
of the Hall bar into close proximity and enhance inter-edge 
scattering amplitudes.  In this paper we consider the momentum
distribution function of such a narrow Hall bar systems.  Using both
Luttinger liquid model and microscopic trial wavefunction approaches,
we find that additional singularities arise which are absent in the limit
of large Hall bars.  The presence of these additional singularities
may play an important role in interpreting transport and 
other experiments in narrow Hall bars.  

The Lagrangian of the chiral Luttinger liquid model\cite{W,KF} for
the edge excitations of a 
Hall bar system occupied by an incompressible state with $\nu =1/M$ is 
\begin{equation}
L=\frac{1}{4\pi\nu} \{ \partial_{x}\phi_{R}(i\partial_{\tau}
+v\partial_{x})\phi_{R}+\partial_{x}\phi_{L}(-i\partial_{\tau}
+v\partial_{x})\phi_{L} \}
\end{equation}
This Lagrangian describes independent excitations at the right(R) and
left(L) edges with oppositely directed velocities of magnitude $v$. 
The fields $\phi_{R}$ and $\phi_{L}$ satisfy the 
commuation rules
\begin{equation}
[\phi_{R}(x),\phi_{R}(x')]=-[\phi_{L}(x),\phi_{L}(x')]=i\pi\nu sgn(x-x')
\end{equation}
The operators for creating an electron and a  
fractionally charged  
particle with charge $\nu e$ at right (R) and left (L)
edges may be written as
$\Psi_{R,L}^{+}(x)=exp(i\phi_{R,L}(x)/\nu)$ and 
$exp(i\phi_{R,L}(x))$, respectively.  
The singularites in $n(k)$ at $ \pm M k_F$ 
follow from this form for the electron creation operator.  
The criteria used to construct these operators are that
they create the appropriate 
amount of charge at the edge and that they have the correct
statistics.  For the electron creation operator there
is abundant numerical evidence\cite{PM} that this {\it ansatz} is 
correct in wide Hall bars.
Although the true microscopic creation operator evidently
contains additional contributions in general, these cannot be 
expressed in terms of the low-energy edge degrees of freedom and will
will not give rise to singularities in $n(k)$. 
We believe that for narrow Hall bars the microscopic electron
creation operator {\it will} have additional low-energy edge contributions.
Additional Boson operators with Fermi statistics and unit charge can  
be created by partitioning the charge between $n$ quasiparticles  
located at the right edge and $l$ quasiparticles located at the 
left edge with $n+l=M$:  
\begin{equation}
\Psi_{n,l}^{\dagger}(x)=exp(in\phi_{R})exp(il\phi_{L})
\end{equation}
where n+l=M.  The propagator of this object may be calculated using the
standard bosonization techniques:\cite{W,KF}
\begin{equation}
<\Psi_{n,l}(x)\Psi_{n,l}^{\dagger}(0)>\sim \frac{exp(i(n-l)k_{F}x)}
{x^{2\Delta_{n,l}}}
\end{equation}
Note that since $n+l=M$ is odd, $p=n-l$ must also be odd. 
By Fourier transforming
$<\Psi_{n,l}(x)\Psi_{n,l}^{\dagger}(0)>$ we see that $n(k)$ has power-law
singularities 
at $\pm pk_{F}$ with the exponents $2\Delta_{p}-1$,
where $p=M, M-2, M-4,...,1$ and 
\begin{equation}
2\Delta_{n,l}=\frac{1}{2}(\frac{1}{\nu}+(n-l)^{2}\nu)
\end{equation}
We see that as $p=n-l$ decreases the exponent $2\Delta_{n,l}$ also decreases.
When the width of the Hall bar greatly exceeds microscopic 
lengths, added electrons will be clearly associated with either 
left or right edges and the contribution of $\Psi_{n,l}^{\dagger}$ 
to the microscopic electron creation operator should become 
small unless $n=0$ or $l=0$. 

To help verify that these new operators do in general contribute to the 
microscopic electron creation operator we consider a specific 
convenient microscopic model of spinless fermions with strong repulsive
short-range interactions in a circularly symmetric confining potential
which confines the electrons to an annulus.  Under appropriate 
circumstances the exact ground state of such a model is given by 
a many-fermion wavefunction of the form
\begin{equation}
\Phi_{M,L}[z]=\prod_{k=1}^{N} z_{k}^{L}\Phi_{M}[z]
\end{equation}
where $L$ is an integer, $N$ is the number of electrons, and 
$\Phi_{M}[z]$ is the Laughlin wavefunction at filling factor 
$\nu=1/M$.  Here a symmetric gauge 
is used so that the single particle 
wavefunctions $\phi_{m}(z)$ are labelled by angular momenta $m$
and are localized around circles of radius $R_m = \ell\sqrt{2(m+1)}$
where $\ell$ is the magnetic length.  In this wavefunction all
the single particle states from $m=L$ to $L+M(N-1)$ are occupied.
Electrons are confined to an annulus with 
inner and outer radii, $R_{1}$ and $R_{2}$, equal to 
$\ell(2(L+1))^{1/2} $ and $\ell(2(L+M(N-1)+1))^{1/2} $ respectively. 
For $L \gg 1$, the width of the annulus $w$ is  
$\approx \ell \frac{M(N-1)-1}{(2L)^{1/2}}$.  For $L=0$ the innner  
radius of the annulus shrinks to zero and the wavefunction 
reduces to the Laughlin wavefunction for the ground electronic
state of a circular electron
droplet whose outer edge can be described by a Luttinger
liquid model.  The wavefunction at finite $L$ can be thought
of as being generated by creating $L$ fractionally
charged quasiholes at the center of the droplet.  For large
$L$ at fixed $N$, the difference between inner and outer radii 
of the electronic annulus becomes small so that the wavefunction
is the ground state wavefunction for a model of a narrow Hall
bar.  In the limit of infinite $L$, the annulus becomes
arbitrarily narrow and the wavefunction becomes equivalent to
that for the Calogero-Sutherland\cite{CS} model of a one-dimensional electron
gas.  Related connections between Laughlin and Calogero-Sutherland
models have been discussed previously.\cite{HR,SF}  We note, for
example, that the exponents characterizing the singularities 
in $n(k)$ predicted by the Luttinger liquid model above are 
identical to those of the of the Calogero-Sutherland one-dimensional
model.\cite{KY}

For this circular geometry, the momentum distribution function of 
a Hall bar maps to the angular momentum distribution\cite{MM} of the 
annulus which is given by 
\begin{equation}
<n_{m}>=\int d^{2}zd^{2}z' \phi_{m}^{*}(z)n(z,z')\phi_{m}(z').
\end{equation}
Here $n(z,z')$ is the one-body density matrix for the ground state
of the system, which we have computed using a 
Monte Carlo method.  
We identify $<n_{m}>$ and $\frac{m-m_{c}}{N/2}$ with $n(k)$ and 
$k/k_{F}$, where $m_{c}=L+M(N-1)/2$ is the angular momentum at the 
center of the annulus.  Then the wavevectors corresponding
to the inner and outer edges are $k/k_{F}=-M(N-1)/N$ 
and $k/k_{F}=M(N-1)/N$.

It is possible to calculate 
$n(k)$ analytically for $k$ near $Mk_{F}$ by following 
Wen\cite{W} and using a plasma analogy.
The total potential energy U of the system is  
\begin{equation}
\frac{-U}{M}=\sum_{i<j}2M\ln| z_{i}-z_{j}|+2L\sum_{k}\ln|z_{k}|
-M\sum_{k}|z_{k}|^{2}/2
\end{equation}
where first, second, and third terms represent, respectively, the mutual
Coulomb interaction energy, the external potential energy
due to the charge L at the origin, and the potential energy due to 
the uniform positive background 
charge. The potential energy of a particle placed at $r>>R_{2}$
is a sum of 
the potential energies  
due to the fixed charges (background charges and electrons in the 
annulus) 
\begin{equation}
V_{fix}(r)=\frac{Mr^{2}}{2\ell^{2}}-\frac{MR^{2}_{2}}{\ell^{2}}ln(r/R_{2})
\end{equation}
and
the screening charge
\begin{equation}
V{sc}(r)=Mln|r-R^{2}_{2}/r|
\end{equation}
The electron density may be evaluated using 
$\rho(r)\propto e^{-(V_{fix}(r)+V{sc}(r))/M}$.  Equating 
$\rho({\bf r})=\sum_{m}n_{m}|\phi_{m}({\bf r})|^{2}$ we
find $n(k)\propto(k\pm Mk_{F})^{M-1}$ near $k=\pm Mk_{F}$.

Fig.1 displays $n(k)$ vs $k/k_{F}$ for M=3 at two values of 
$w=1.24\ell$ and $2.52\ell$. To extract 
the exponent at $k=\pm k_{F}$ we use a scaling ansatz for the
momentum distribution function of a finite sample with length R
\begin{equation}
n(k,R)\propto R^{-\alpha}f(|k-k_{F}|R)
\end{equation}
where $R=2\pi\ell\sqrt{2(m_{c}+1)}$, $\mid k-k_{F} \mid /k_{F} \ll1$, 
and the scaling function 
$f(x)\sim x^{\alpha}$ for large x. The difference between $n(k)$ 
at two values of $k$ adjacent to $k_{F}$ is
\begin{equation}
\Delta n=n(k_{F}^{+},R)-n(k_{F}^{-},R)=g(w)R^{-\alpha}
\end{equation}
where $k_{F}^{+}/k_{F}$ and 
$k_{F}^{-}/k_{F}$ are $1+1/N$ and $1-1/N$.
Analysing our numerical data accordingly we find $\alpha=0.6\pm0.1$.  
This estimate is consistent
with the exact result $2/3$.  We have also investigated
the width dependence $g(w)$ and find that it is approximately 
exponential: $g(w)\sim exp(-w/a)$ with $a$ about $1.6\ell$.
The exponent
at $k=\pm3k_{F}$ is numerically verified to be 2, in agreement with the 
analytical result.  In Fig.2 we plot n(k) vs $(\frac{k}{k_{F}}+3)^{2.4}$
for M=5. From the linear dependence we deduce that the relevant
exponent is 2.4, in agreement with the Luttinger liquid model predictions. 

In summary, we conclude on the basis of both microscopic many-particle
wavefunctions and Luttinger liquid model that in a narrow Hall bar 
$n(k)$ has singularities at 
$k=\pm pk_{F}$, where $p=M,M-2,M-4,...,1$
and that the magnitude of the corresponding
exponents decreases with decreasing $p$.
For both calculations we have worked with models in 
which inter-edge interactions are absent so that the 
exponents are independent of the Hall bar width, while
the amplitudes of the singularities vanish exponentially
with $w$ for $p \ne M$.  These new and 
stronger singularities will be important\cite{F}  
in the interpretation of experimental searches
for Luttinger liquid behavior in narrow Hall bars. 

E. Yang thanks S. J. Rey for useful conversations.  This work was 
supported in part by NON DIRECTED RESEARCH FUND, 
Korea Research Foundation, and in part by the National
Science Foundation under grants DMR-9416906, DMR-9400142, DMR-9528578 and
PHY-9407194.

\begin{figure}
\caption{$n(k)$ is plotted vs $k/k_{F}$ for $w=1.24\ell$ and 
$2.52\ell$.  Data collapse is seen for each value of w. The plotting symbols
are triangles, circles, squares, diamonds, triangle lefts, 
and triangle downs and their $(N,M,L)$ values are (3,3,8), 
(4,3,21), (5,3,40), 
(6,3,65), (6,3,10), and (7,3,16), respectively. 
\label{fig1}}
\end{figure}

\begin{figure}
\caption{$n(k)$ is plotted vs $(\frac{k}{k_{F}}+3)^{2.4}$ 
for (N,M,L)= (9,5,218).  The linear dependence
implies that the relevant exponent is 2.4.                 
\label{fig2}}
\end{figure}

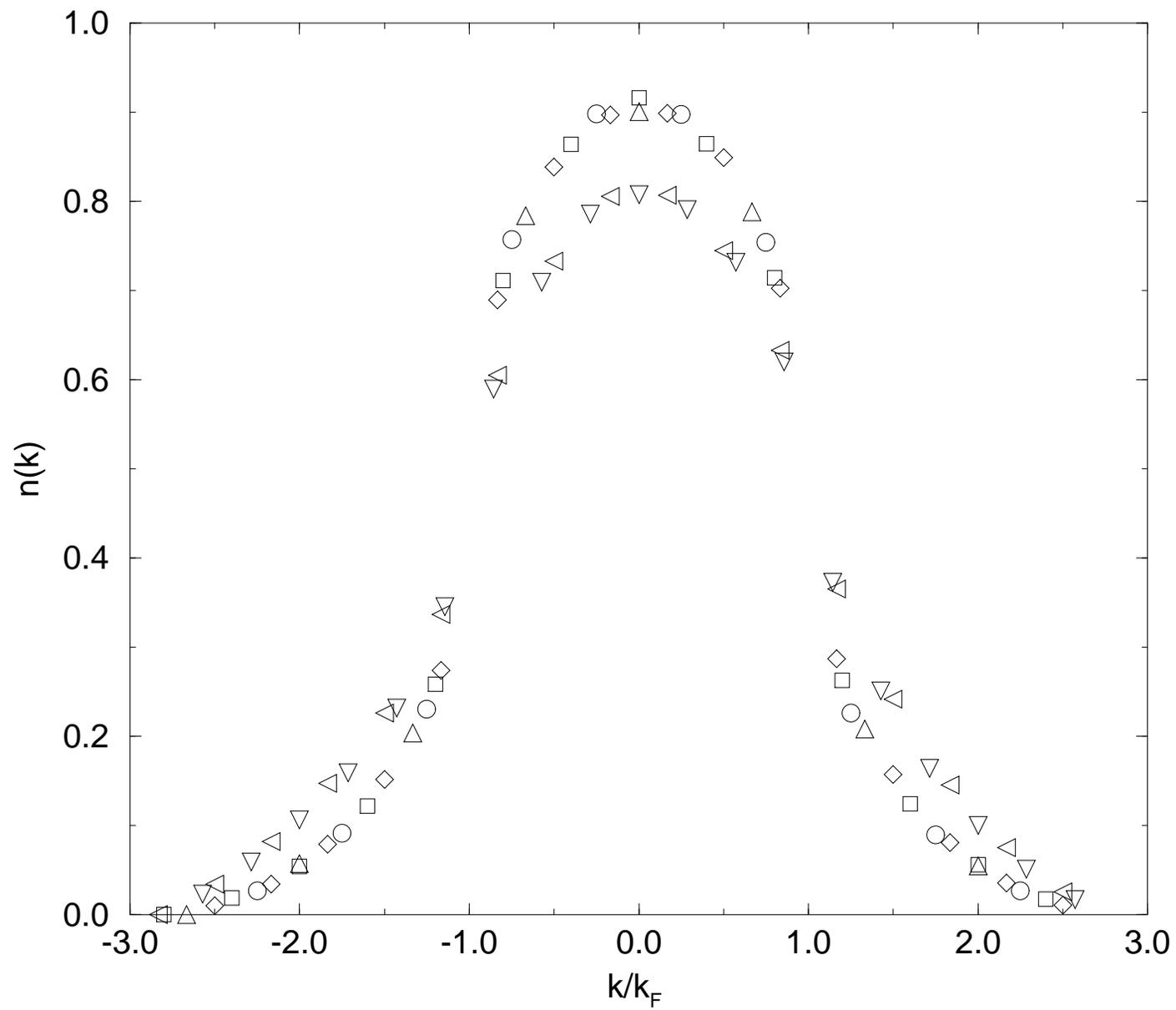
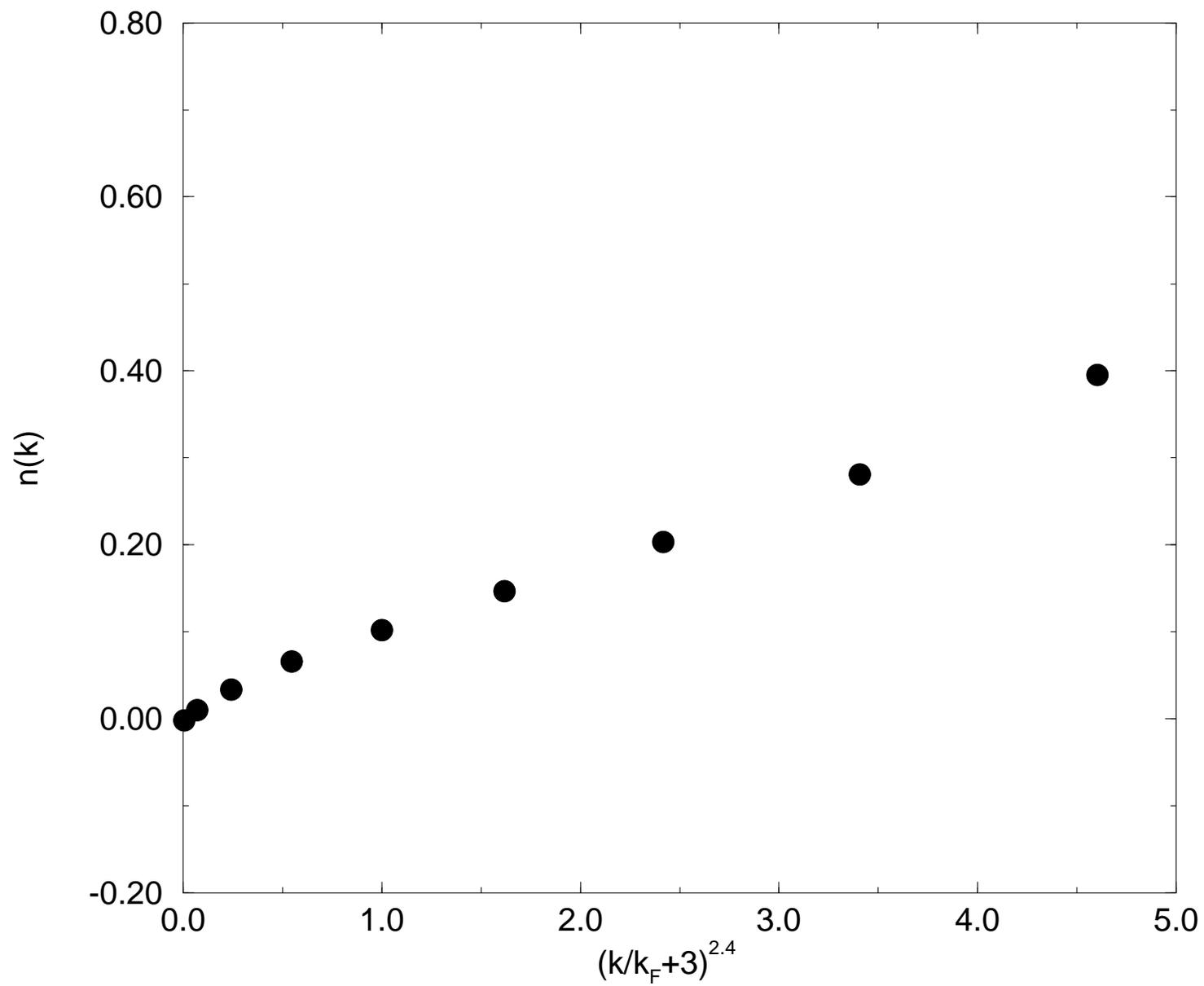
\end{document}